\documentclass{amsart}
\usepackage{a4,amsmath,amssymb,latexsym,enumerate}
\newcommand{\supp}{{\mathop{\mathrm{supp}}}}

\begin{document}
\title{Spectral properties of Anderson-percolation Hamiltonians}

\author[I.~Veseli\'c]{Ivan Veseli\'c}

\address{Emmy-Noether Programme of the DFG  \& Fakult\"at f\"ur Mathematik, D-09107 TU Chemnitz, Germany}

\urladdr{www.tu-chemnitz.de/mathematik/schroedinger/members.php}
\thanks{\copyright 2006 by the author. Faithful reproduction of this article is permitted for non-commercial purposes.}

\maketitle

This is a slightly enlarged and corrected version of a contribution to the Oberwolfach Report \cite{IV-7}.
We discuss results on spectral properties of 
Laplacians on percolation graphs and more general Anderson-percolation Hamiltonians. 
The results are mostly taken
from the papers \cite{IV-1,IV-2,IV-3}, to which we refer for further
references, generalisations of the results presented here, and more details.
\smallskip

In this report we will for simplicity restrict ourselves to the following situation.
We consider the graph with vertex set $\mathbb{Z}^d$, where two of the vertices
are joined by an edge if their $\ell^1$ distance is equal to one.
The group $\Gamma:=\mathbb{Z}^d$ is acting on the graph by translations. 
Denote by $\{\Lambda_L\}_{L\in 2 \mathbb{N}}$ the exhaustion of $\mathbb{Z}^d$ by cubes 
$\Lambda_L := \mathbb{Z}^d \, \cap \, [-L/2,L/2]^d$.

We construct a probability space $(\Omega, \mathcal{A}, \mathbb{P})$ associated
to percolation on $\mathbb{Z}^d$. Let $\Omega= \times_{x\in \mathbb{Z}^d} [0, \infty]$
be equipped with the $\sigma$-algebra $\mathcal{A}$ generated by finite dimensional cylinders sets.
Denote by $\mathbb{P}$ a probability measure on $\Omega$ and assume that the measurable shift transformations
$ \tau_\gamma\colon \Omega \to \Omega$, $(\tau_\gamma \omega)_x = \omega_{x-\gamma}$
are measure preserving. Moreover, let the family $\tau_\gamma , \gamma \in \Gamma$
act ergodically on $\Omega$.
By the properties of $\tau_\gamma , \gamma \in \Gamma$
the stochastic field $q\colon \Omega\times \mathbb{Z}^d \to [0,\infty]$
given by $q(\omega,x)=\omega_x, x \in \mathbb{Z}^d$ is \emph{stationary} or \emph{equivariant},
i.e.{} $q(\tau_\gamma\omega,x)=q(\omega,x-\gamma) $.
The mathematical expectation associated to the probability $\mathbb{P}$ will be denoted by $\mathbb{E}$
and the distribution measure of $q_0$ by $\mu$.
\smallskip

Define for $\omega\in \Omega$ the random vertex set 
$X(\omega) := \{ x \in  \mathbb{Z}^d \mid q(\omega,x) <\infty\}$ and denote by the same symbol the 
induced subgraph of $\mathbb{Z}^d$. In other words, if $q(\omega,x)$ is infinite, 
we delete the vertex $x$ together with all incident edges, 
otherwise we retain it in the graph. 
For each $\omega$ let $A_\omega\colon \ell^2(X(\omega))\to \ell^2(X(\omega))$ 
be the adjacency operator of the graph $X(\omega)$. 
Define now a selfadjoint Anderson-percolation Hamiltonian by
$D(\omega):=\{f \in \ell^2(X(\omega))\mid \sum_{x\in X(\omega)} |q(\omega,x) f(x)|^2<\infty\}$
and
\begin{equation*}
H_\omega := A_\omega + q(\omega,\cdot)  \colon D(\omega)\to D(\omega)
\end{equation*}
where $(q(\omega,\cdot)\psi)(x) = q(\omega,x)\psi(x)$ is a random potential.
Note that $c(\omega):= \{f \in \ell^2(X(\omega))\mid \supp \, f \text{ is finite}\}$
is an operator core for $H_\omega$, for all $\omega$.
In the sequel we may or may not assume
\begin{equation}
\label{nopotential}
\supp \ \mu = \{0, \infty\}
\end{equation}
\begin{equation}
\label{iid}
q(\cdot,x) , x \in\mathbb{Z}^d \quad \text{ are i.i.d.~random variables }
\end{equation}

Let us consider two examples of random potentials.
Set $V(\omega,x)= 2d -\mathrm{deg}_{X(\omega)} (x) \ge 0$. Then the two operators $A_\omega \pm V(\omega,\cdot)$
are sometimes called Neumann, respectively Dirichlet Laplacian on $X(\omega)$ (up to a constant). 
The fixed sign of the potential $V$ is useful for Dirichlet-Neumann-bracketing estimates.
\smallskip

\textbf{Remark.}
More generally, Anderson-percolation Hamiltonians can be defined on covering graphs 
with a free, coompact group action. For certain results it is required that the group is amenable.
The structure of the stochastic process $q$ generalises in an obvious manner to this setting.
It is possible to carry out a similar analysis on bond or mixed percolation graphs.
Moreover, rather than considering the adjacency operator, one can derive similar results for 
equivariant, hermitian, finite hopping range operators. In this generality, some of our results are new 
even in the periodic case when $\Omega$ contains only one element. 
\smallskip

Denote by $\sigma_{disc}, \sigma_{ess}, \sigma_{ac}, \sigma_{sc}, \sigma_{pp}$
the discrete, essential, absolutely continuous, singular continuous, and pure point part of the spectrum, 
and by  $\sigma_{fin}$ the set of eigenvalues which posses an eigenfunction with finite support. 
For each $\Lambda_L$ denote by $H_\omega^L$ the truncation of $H_\omega$ to $\ell(\Lambda_L(\omega))$, $\Lambda_L(\omega):=
\Lambda_L\cap X(\omega)$ and the associated normalised eigenvalue counting function by
$ N_\omega^L({E}):= \frac{1}{L^d}\mathrm{Tr} [ \chi_{]-\infty,E]}(H_\omega^L)]$.
The distribution function, given by an averaged trace per unit volume, 
$E \mapsto N(E) =\mathbb{E} \left \{ \langle \delta_0, \chi_{]-\infty,E]}(H_\omega) \delta_0  \rangle\right \} $
is called \emph{integrated density of states} (IDS). Denote by $\nu$ the measure on $\mathbb{R}$ associated to $N$.
The following theorem establishes the non-randomness of basic spectral quantities and
a relation between the IDS and the spectrum.
\smallskip

\textbf{Theorem 1.}
There exists an $\Omega' \subset \Omega$ of full measure and subsets of the
real numbers $\Sigma$ and $ \Sigma_\bullet$, where $\bullet \in\{disc, ess, ac, sc, pp, fin\}$,
such that for all $\omega\in \Omega'$
\[
\sigma(H_\omega)=\Sigma \quad \text{ and } \quad \sigma_\bullet (H_\omega)= \Sigma_\bullet
\quad \text{ for any $\bullet = disc, ess, ac, sc, pp, fin$ } 
\]
and
\begin{equation}
\label{t-d-IDS}
\lim_{L\to \infty} N_\omega^L (E) = N(E) \quad \text {at all continuity points of $N$}
\end{equation}\noindent
Moreover, $\Sigma_{disc}=\emptyset$ and  $\Sigma = \mathrm{supp}\, \nu$.
Under assumption \eqref{nopotential}  the convergence \eqref{t-d-IDS} holds for all $E \in \mathbb{R}$.
\smallskip

There is a probability threshold $p_c$ such that if $\mu (\{\infty\})< 1-p_c $ and \eqref{iid} holds
\begin{equation}
\label{percolates}
\text{an infinite component } X^\infty(\omega) \text{ of the graph } X(\omega) \text{ exists a.s.}
\end{equation}
In this case one can consider the adjacency operators $A_\omega^\infty$ on $X^\infty(\omega)$
and the analogous statements to the above theorem hold.
We denote the associated quantities using a superscript $^\infty$, e.g. $N^\infty$ for the IDS of $A_\omega^\infty$ .
\smallskip

\textbf{Theorem 2.} 
\begin{enumerate}[(i)]
\item
$\Sigma_{fin}= \mathrm{supp} \ \nu_{pp}$. 
\item
Denote by  $A^G$ the adjacency operator of the subgraph of $\mathbb{Z}^d$ induced by $G$.
If \eqref{nopotential} and \eqref{iid} hold, then
\[
\Sigma_{fin}=\tilde\Sigma :=  \{ E \in \mathbb{R} \mid \exists \text{ finite } G\subset \mathbb{Z}^d \text{ and } f \in \ell^2(G)
\text{ such that } H^G f=Ef\} 
\]
\item
If \eqref{percolates} holds we have 
$\Sigma_{fin}^\infty  = \mathrm{supp} \ \nu_{pp}^\infty$. 
\item 
\label{i-adjec-op}
If \eqref{percolates} and \eqref{iid} hold then $\Sigma_{fin}^\infty  = \Sigma_{fin}$. 
\end{enumerate}

Note that the set $\tilde\Sigma$ is dense in $[-2d, 2d]$ and consists of algebraic integers.\\

Statement (i) of Theorem 1 describes the location of the jumps of the IDS in terms of 
finitely supported eigenfunctions. In fact, it is possible to describe the size of the jumps in this way, too.
A special case of the result in \cite{IV-6} is 
\smallskip

\textbf{Proposition}\\
\noindent
Let $E \in \mathbb{R}$. Then there exists an $\Omega' \subset \Omega$ of full measure
such that for all $\omega \in \Omega'$
\[
\{ f\in D(\omega) \mid H_\omega f = Ef \}
=
\overline{\{ f\in D(\omega) \mid H_\omega f = Ef, \supp \ f \text{ is finite} \}}
\]

The next theorem provides  estimates in the spirit of Wegner and Delyon/Souillard 
for Anderson-percolation Hamiltonians.
\smallskip

\textbf{Theorem 3.} Assume \eqref{iid}, then:
\begin{enumerate}[(i)]
\item
$\Sigma= [-2d,+2d] + \mathrm{supp}\ \mu|_{\mathbb{R}}$. 
\item
$\Sigma_{fin} \supset \tilde \Sigma + \mathrm{supp}\ \mu_{pp}|_{\mathbb{R}}$
\item
If $\mu=\mu_c +(1-p)\delta_\infty$, i.e.~$\mu$ has no atoms at finite values, 
then the IDS of $H_\omega$ is continuous.
\item 
Assume  that for $a,b\in \mathbb{R}$ the measure $\mu$ is absolutely continuous on the interval $]a-2d, b+2d[$, 
i.e.~$\mu|_{]a-2d, b+2d[}(dx)=f(x)dx$, and that $f\in L^\infty$. Then, for every interval $I$ with  $\mathrm{dist}(I,]a,b[^c)\ge \delta>0$ we have
\begin{equation}
\label{e-WE}
\mathbb{E} \{\mathrm{Tr} \, [\chi_I(H_\omega^L)] \} \le C \, |I| \, L^d
\end{equation}
where $C= 2^{d+2}\, \left(\frac{b-a+4d+1}{\delta}\right)^2   \frac{\|f\|_\infty }{\mu (]a-2d, b+2d[)} $.
\end{enumerate}

It follows that the constant $C$ in \eqref{e-WE} is an upper bound on the \emph{density of states}, i.e.~$
\frac{dN(E)}{dE} \le C \text{ for all }  E \in ]a,b[$.
\bigskip

If \eqref{iid} holds and 
$\mu=\frac{1}{3}\chi_{[0,1]} +\frac{1}{3}\delta_{4d}+\frac{1}{3}\delta_\infty$,
 the Wegner estimate \eqref{e-WE} is valid for all energies  $E <2d$. Moreover, by \cite{IV-4},
Lifshitz asymptotics hold at $\min \Sigma= -2d$. 
Thus one can prove using a multiscale analysis, see e.g.~\cite{IV-5}, 
that for some $\alpha \in\, ]0,1[$ the spectrum of $H_\omega$ in $U = [-2d,-2d+\alpha]$ 
has no continuous component and consists of a dense set of eigenvalues,
whose eigenfunctions decay exponentially in space almost surely.

Denote by $\sigma_\epsilon(H_\omega)$ the set of eigenvalues of $H_\omega$. 
Then 
\begin{equation}
\Sigma_{fin} \subsetneqq \sigma_\epsilon(H_\omega) \subsetneqq \Sigma_{pp} 
\quad\text{ for almost all } \omega
\end{equation}
Here the sets $\Sigma_{fin}$ and $\Sigma_{pp}$ are almost surely non-random, and
$\Sigma_{fin}\supset \tilde \Sigma + 4d$. 
The set $\sigma_\epsilon(H_\omega) \, \cap \, U$ is non-empty almost surely and disjoint to $\tilde \Sigma + 4d$.
By the Wegner estimate the IDS is absolutely continuous in $U$, therefore 
for any $E \in U$ the probability $\mathbb{P}\{E \text{ in an eigenvalue of } H_\omega\}$ vanishes
and $U$ is even disjoint to $\Sigma_{fin}$. 
Thus we have a highly fluctuating set $\sigma_\epsilon(H_\omega)$ strictly sendwiched 
between two almost surely non-random sets $\Sigma_{fin}$ and $\Sigma_{pp}$.


\begin{thebibliography}{99}
\bibitem{IV-4}  M.~Biskup, W.~K{\"o}nig,
\newblock Long-time tails in the parabolic {A}nderson model with bounded potential
\newblock Ann. Probab. 29(2):636--682, 2001.
\newblock http://www.arXiv.org/math-ph/0004014.

\bibitem{IV-7} J.~Dodziuk, D.~Lenz, N.~Peyerimhoff, T.~Schick, I.~Veseli\'c, Editors.
\newblock $L^2$-Spectral Invariants and the Integrated Density of States.
\newblock Oberwolfach Reports. \textbf{3}(1):511--552, 2006.
\newblock http://www.mfo.de/programme/schedule/2006/08b/OWR\_2006\_09.pdf


\bibitem{IV-5} H.~von Dreifus and A.~Klein.
\newblock A new proof of localization in the {Anderson} tight binding model.
\newblock {\em Commun. Math. Phys.}, 124:285--299, 1989.

\bibitem{IV-3} D.~Lenz, N.~Peyerimhoff, and I.~Veseli\'c.
\newblock Von {Neumann} algebras, groupoids and the integrated density of
  states.
\newblock http://www.arXiv.org/math-ph/0203026.

\bibitem{IV-6} D.~Lenz, I.~Veseli{\'c}.
\newblock Hamiltonians on discrete structures: jumps of the integrated density of states.
\newblock in preparation.

\bibitem{IV-1} I.~Veseli{\'c}.
\newblock Quantum site percolation on amenable graphs.
\newblock In {\em Proceedings of the Conference on Applied Mathematics and
  Scientific Computing}, pages 317--328, Dordrecht, 2005. Springer.
\newblock http://arXiv.org/math-ph/0308041.

\bibitem{IV-2}  I.~Veseli\'c.
\newblock Spectral analysis of percolation {Hamiltonians}.
\newblock {\em Math.~Ann.}, 331(4):841--865, 2005.
\newblock http://arXiv.org/math-ph/0405006.
\end{thebibliography}
\end{document}